\documentclass[12pt]{article}
\usepackage{amssymb,amsmath} 
\usepackage{graphicx}
\graphicspath{{graphics/}}
\usepackage{latexsym}
\usepackage{subfig}  
\usepackage{booktabs,setspace,caption}  
\usepackage{braket}
\usepackage{wasysym,verbatim}
\usepackage[numbers,sort&compress]{natbib}
\usepackage{textcomp}
\usepackage[utf8]{inputenc}

\usepackage{mathrsfs}
\usepackage[inner=2.9cm,outer=2cm]{geometry}
\usepackage[toc]{appendix}
\usepackage{color,soul} 
\usepackage{datetime}
\usepackage[
      colorlinks=true,
      linkcolor=darkblue,  
      urlcolor=black,    
      filecolor=blue,     
      citecolor=blue, 
      linktocpage=true,
      pdfstartview=FitV,
      bookmarksopen=true    
      ]{hyperref}

\definecolor{darkblue}{rgb}{0.2, 0, 0.8}
\definecolor{darkgreen}{rgb}{0.2, 0.71, 0}

\newcommand{\eq}{\begin{equation}}
\newcommand{\qe}{\end{equation}}

\numberwithin{equation}{section}

\newenvironment{changemargin}[2]{%
\begin{list}{}{%
\setlength{\topsep}{0pt}%
\setlength{\leftmargin}{#1}%
\setlength{\rightmargin}{#2}%
\setlength{\listparindent}{\parindent}%
\setlength{\itemindent}{\parindent}%
\setlength{\parsep}{\parskip}%
}%
\item[]}{\end{list}}

\begin{document}  


\begin{titlepage}

\begin{flushright}
{\tt \small{IFT-UAM/CSIC-19-125}} \\
{\tt \small{IFUM-1079-FT}} \\
\end{flushright}

\vspace*{1.2cm}

\begin{center}
{\Large {\bf{On the extremality bound of stringy black holes} }} \\

\vspace*{1.2cm}
\renewcommand{\thefootnote}{\alph{footnote}}
{\sl\large Pablo A. Cano$^{\text{\quarternote }}$, Tom\'as Ort\'in$^{\text{\quarternote }}$ and Pedro F.~Ram\'{\i}rez$^{\text{\eighthnote}}$ }\footnotetext{{pablo.cano[at]uam.es, tomas.ortin[at]csic.es, ramirez.pedro[at]mi.infn.it
}}
\bigskip

$^{\text{\quarternote}}$Instituto de Física Teórica UAM/CSIC,\\
C/ Nicolás Cabrera, 13-15, C.U. Cantoblanco, 28049 Madrid, Spain\\

\bigskip

$^{\text{\eighthnote}}$INFN, Sezione di Milano,\\
Via Celoria 16, I-20133 Milano\\

\setcounter{footnote}{0}
\renewcommand{\thefootnote}{\arabic{footnote}}

\bigskip

\bigskip

\end{center}

\vspace*{0.1cm}

\begin{abstract}  
\begin{changemargin}{-0.95cm}{-0.95cm}
\noindent  
A mild version of the weak gravity conjecture (WGC) states that extremal black
holes have charge-to-mass ratio larger or equal than one when higher-curvature
interactions are taken into account. Since these corrections become more
relevant in the low-mass regime, this would allow the decay of extremal black
holes in terms of energy and charge conservation. Evidence in this direction
has been mainly given in the context of corrections to Einstein-Maxwell
theory. Here we compute corrections to the charge-to-mass ratio of some dyonic
extremal black holes explicitly embedded in the heterotic string effective
theory. We find that modifications of the extremality bound depend on the
solution considered, with the charge-to-mass ratio remaining unchanged or
deviating positively from one. Additionally, we observe that the introduction
of the higher-curvature terms increases the Wald entropy in all cases
considered, whose variation does not seem to be correlated with the
charge-to-mass ratio, contrary to the situation in Einstein-Maxwell
theory. 
\end{changemargin}
\end{abstract} 

\end{titlepage}

\setcounter{tocdepth}{2}
{\small
\setlength\parskip{-0.5mm} 
\noindent\rule{15.7cm}{0.4pt}
\tableofcontents
\vspace{0.6cm}
\noindent\rule{15.7cm}{0.4pt}
}

\section{Introduction}

Black holes have played, and continue to play, a central role in fundamental
aspects of physics. Cutting-edge advances in the understanding of quantum
gravitational aspects of string theory have been possible thanks to their
study \cite{Kallosh:1992ii, Ferrara:1995ih, Strominger:1996sh, Callan:1996dv,
  Ferrara:1996dd, Ferrara:1996um, Sen:2005wa, Mathur:2005zp,
  Papadodimas:2013jku}. In recent years some attention has been dedicated to
the problem of whether extremal black holes, those with vanishing temperature,
should decay or remain as stable states. The question arises because, in many
situations, black hole remnants are regarded as problematic. For example, it
can be argued that a theory of gravity with global symmetries contains an
infinite number of remnants below a certain mass scale, which is usually
considered to be inconsistent \cite{Banks:2010zn, Susskind:1995da}. On the
other hand, stable black holes of theories with local symmetries are not
infinitely degenerated (under a certain mass scale), so it is more difficult
to produce arguments against their existence. Still, they are associated to an
infinite tower of charged states not protected by a symmetry principle, and it
was conjectured in \cite{ArkaniHamed:2006dz} that a finiteness criterion
should be applied such that those are unstable.

Hence, we are led to consider the decay of extremal black holes. The simplest
of those is the Reissner-Nordstr\"om one, which has $M=Q$ in appropriate
units. It is immediate to see that decay of this black hole in two separated
states with $q_1+q_2=Q$, $m_1+m_2 \leq M$ requires $q_i/m_i \geq 1$ for at
least one of them\footnote{This is a necessary but not sufficient condition
  for the process to occur, which seems thermodynamically disfavored. Since
  these black holes have zero temperature, standard Hawking radiation does not
  take place.}. These bounds can be saturated in the special case in which
there is no binding energy between the products, while an strict inequality 
is expected in generic situations.

The weak gravity conjecture proposes that the spectrum of a quantum theory of
gravity must be such that extremal black holes can decay, as far as energy and
charge conservation are concerned. There are two possible scenarios compatible
with the conjecture. Under the rather reasonable assumption that the decay
would occur through the emission of a \emph{light particle} by the black hole,
\emph{i.e.} $m_1 >> m_2$, the consequences of having $q_2/m_2>1$ are stronger
than those of the complementary scenario $q_1/m_1>1$ in the following
sense. In terms of the former, the WCG becomes a sharp tool that can be useful
to discern if an effective theory belongs or not to the swampland
\cite{Vafa:2005ui, ArkaniHamed:2006dz}. Examples of applications of the
conjecture in this direction can be found in \cite{Palti:2019pca} and
references therein. On the other hand, the latter possibility corresponds to a
milder version of the conjecture that would just provide information about
states far heavier than the Planck mass, and hence it is less useful for the
swampland program. Of course, the two scenarios are not mutually exclusive,
and it is conceivable that they might be related \cite{Aalsma:2019ryi}.

The problem of computing corrections to the extremal charge-to-mass bound has
been considered before by other authors in several frameworks. To date, the
Reissner-Nordstr\"om black hole of Einstein-Maxwell theory supplemented by
higher-derivative terms is arguably the system which is best understood in
that respect. An explicit computation for that system that gives the
corrections to the ratio in terms of the value of the coefficients of the
higher-derivative terms was presented in \cite{Kats:2006xp} (see also \cite{Loges:2019jzs}). Subsequent works
have proposed that, demanding analyticity of scattering amplitudes, unitarity
and causality constrain these coefficients such that there is a positive
deviation of the charge-to-mass ratio of extremal black holes
\cite{Cheung:2014ega, Hamada:2018dde, Bellazzini:2019xts}. Likewise, it has
been proposed that this can be related to the positivity of the corrections to
the entropy of the black hole induced by higher-derivative operators
\cite{Cheung:2018cwt, Cheung:2019cwi,Goon:2019faz}. The study of black holes in
Einstein-Maxwell theory is well justified and interesting, as it provides a
relatively simple arena to explore this question while making contact with
dominant interactions in real world experiences. On the other hand, since the
problem at hands is intimately related to quantum gravity, it is important to
ask if the positive character of the deviation is displayed by models with an
explicit string theory embedding.

Effective gravitational theories derived from string theory usually contain,
besides vectors, scalars. Einstein-Maxwell-Dilaton (EMD) theory arises as a
natural truncation of the effective theories of different string models. For
instance, it appeared as a truncation of $N=4,d=4$ supergravtiy\footnote{This
  is the effective theory of the Heterotic Superstring compactified on $T^{6}$} 
in \cite{Gibbons:1982ih}, where the first EMD black-hole solutions where
found. These solutions were later rederived and studied in
\cite{Gibbons:1987ps, Garfinkle:1990qj}.\footnote{This solution is usually
  referred as the GHS black hole.} The Kaluza-Klein theory obtained by
compactifying the 5-dimensional Einstein-Hilbert action on a circle also
provides another particular example of EMD theory with the Kaluza-Klein scalar
playing the role of dilaton field. Different instances of EMD theory are
distinguished by the different couplings of the dilaton to the vector field
kinetic term in the action.

Although EMD theory looks very similar to Einstein-Maxwell (EM), there are
important differences. In particular, the purely electrically or magnetically
charged Reissner-Nordstr\"om black hole with constant scalar is not a solution
of the equations of motion, as the vector field acts as a source for the
dilaton. Only when the black hole is dyonic with equal magnetic and electric
charges the source term in the dilaton equation vanishes and the dilaton can
take a constant value. For generic values of the electric and magnetic
charges, one gets charged black holes with a non-trivial scalar field.

In the extremal limit, these black holes are regular,
except in the purely electric and purely magnetic cases. Corrections to the charge-to-mass ratio in one of these singular
cases have been studied in \cite{Kats:2006xp}, although, due to the
singularities, this is not a good ground to discuss stability of extremal
black holes. Still, it is worth mentioning that the correction has again a
positive character, which gives some support to the mild WGC.

However, the positive deviation of the charge-to-mass ratio cannot be general
in string theory. Supersymmetric black holes are special. When they are
regular they necessarily carry several charges and
their mass is given by a linear combination of them with moduli-dependent
coefficients. Typically, the scalars are not constant, but their value at the
horizon is fixed in terms of the charges due to the attractor mechanism
\cite{Ferrara:1995ih, Strominger:1996kf, Ferrara:1996dd, Ferrara:1996um,
  Ferrara:1997tw, Ferrara:2006em}. The linear relation between mass and charge
is a salient feature of supersymmetric systems and, as supersymmetric black
holes are extremal \cite{Kallosh:1992ii}, the charge-to-mass ratio can be
expected to remain unmodified by higher-derivative corrections. This has been
recently shown to be the case in three- and four-charge heterotic black holes
\cite{Cano:2018qev, Cano:2018brq} in, respectively, five and four
dimensions.\footnote{In those articles, it was noted that the relation between mass
and the number of fundamental constituents of the black hole is modified
by $\alpha'$-corrections. However, the relation between mass and asymptotic charges 
remains unchanged.}
 Nevertheless, one should notice that these configurations
correspond to a bound state of a (large) number of fundamental objects without
binding energy. This means that, speaking in terms of energy and charge
conservation, the decay of supersymmetric black holes is possible. Another
family of configurations of $\mathcal{N}\geq2$ four-dimensional supergravity
for which no corrections to the ratio have been observed (even at one-loop
quantum level and non-supersymmetric solutions) was recently described
\cite{Charles:2019qqt}. These black holes are obtained through a particular
embedding of dyonic solutions of Einstein-Maxwell theory for which the
solution is claimed to not receive corrections at all \cite{Charles:2017dbr}.

In this article we compute explicitly the first-order $\alpha'$ (fourth order
in derivatives) corrections to the charge-to-mass ratio of the extremal
Reissner-Nordstr\"om black hole embedded in heterotic string theory in several
ways. All the embeddings that we will consider here are dyonic (so the dilaton
vanishes at lowest order in $\alpha'$) and non-supersymmetric (so there is a
chance of having non-vanishing $\alpha'$ corrections).  To the best of our
knowledge, these provide the first examples in which such a computation has
been made using a explicit embedding of the black-hole solutions in a
superstring theory whose first-order in $\alpha'$ corrections are explicitly
known in detail.

We start in Section~\ref{sec-family}
with a description of the zeroth-order solutions we start from. They are
2-vector dyonic, extremal Reissner-Nordstr\"om black holes, although we take
only one of the charges to be independent for simplicity. Depending on the
election of the relative signs of the charges, there are two families of
solutions that can be considered, for which the consequences of including the
higher-curvature corrections are different. In one case the charge-to-mass
ratio of the solution remains unchanged, while in the other it deviates
positively from one. On the other hand, the Wald entropy of both solutions is
equal and differs from the expression of the zeroth-order system. Our results are discussed in Section~\ref{sec-discussion}. 

In the different appendices we include the details about the different computations that we have performed. The effective field theory of the heterotic string at first order in $\alpha'$ is described in Appendix~\ref{sec:theory}. The equations of motion evaluated for the spherically-symmetric ansatz can be found in Appendix~\ref{sec-eom} and the calculation of the Wald entropy is addressed in Appendix~\ref{sec:wald}.
In Appendix~\ref{sec:case3} we compute the corrections for a solution different to the ones considered in Section~\ref{sec-family}, which has some peculiar features.

\section{A family of extremal black holes}
\label{sec-family}

Let us consider the following field configuration of heterotic superstring theory, whose perturbative action and equations of motion are briefly reviewed in Appendix \ref{sec:theory},
\begin{align} \nonumber 
d\hat{s}=&e^{2( \phi- \phi_{\infty})}ds^2-c^2(dz+V/c_{\infty})^2-dy^idy^i\, , \\ \nonumber
\hat H=&F\wedge (c_{\infty}dz+V)+H\, ,\\ \label{eq:comapct}
e^{-2\hat\phi}=&\frac{1}{c}e^{-2\phi} \, ,
\end{align}
where $ds^2$ is the four-dimensional metric in the Einstein frame, $F$ is a 2-form, $V$ is a Kaluza-Klein vector, $c$ is a KK scalar, $H$ is a 3-form, and $\phi$ is the four-dimensional dilaton. In addition, $c_{\infty}$ and $ \phi_{\infty}$ are the asymptotic values of $c$ and $\phi$. 
These are effective four-dimensional fields, while hatted objects represent ten-dimensional fields of the heterotic theory.
This ansatz corresponds to the compactification of heterotic superstring theory on $\mathbb{S}^1_z\times \mathbb{T}^5$, where we truncate all the fields that have indices on $\mathbb{T}^5$, while the KK reduction on $\mathbb{S}^1_z$ is general. The coordinates parametrizing the compact space $z$ and $y^{i}$, $i=1,2,3,4,5$ have all period $2\pi \ell_s$. 

At the supergravity level (zeroth-order in $\alpha'$), this gives rise to the following four-dimensional effective theory
\begin{equation}\label{eq:hetred}
\begin{aligned}
S=\frac{1}{16\pi G_N^{(4)}}\int d^4x\sqrt{|g|}\Bigg\{&R+2(\partial\phi)^2+\frac{(\partial c)^2}{c^2}+\frac{e^{-4(\phi-\phi_{\infty})}}{2\cdot 3!}H^2\\
&+\frac{e^{-2(\phi-\phi_{\infty})}}{4}\left(G^2+\frac{c_{\infty}^2}{c^2}F^2\right)\Bigg\}\, ,
\end{aligned}
\end{equation}
where $G=dV$ and, at this order, $F=dA$. The four-dimensional Newton's constant is given by
\begin{equation}
G_N^{(4)}=\frac{G_N^{(10)}}{c_{\infty}(2\pi\ell_s)^6}\, .
\end{equation}
Also, the 3-form satisfies the Bianchi identity
\begin{equation}
dH=-F\wedge G\, ,
\end{equation}
and using it we can dualize $H$ into a scalar field.  From the effective four-dimensional action \eqref{eq:hetred} one sees that, at the supergravity level, one could truncate $V$, $H$ and $c$. This would simplify the system to the Einstein-Maxwell-Dilaton model. However, it turns out that this is inconsistent once $\alpha'$ corrections are taken into account, as those introduce non-trivial couplings between these fields. In other words, higher-derivative corrections to the Einstein-Maxwell-Dilaton effective model in the context of string theory may require the activation of additional fields. This is a well-known but often forgotten fact \cite{Campbell:1991kz, RNCORR}.
 
\subsection{Supergravity zeroth-order solution}

A generalized version of the extremal Reissner-Nordstrom black hole can be a solution of the theory (\ref{eq:hetred}) if we allow for dyonic vectors. Let us consider the following configuration, 
\begin{eqnarray} \nonumber
ds^2&=&\left(1+\frac{Q}{r}\right)^{-2}dt^2-\left(1+\frac{Q}{r}\right)^2\left(dr^2+r^2d\Omega_{(2)}^2\right)\, ,\\ \nonumber
A&=&\frac{2q_A}{(r+Q)}dt-2p_A\cos\theta d\varphi\, ,\\ \nonumber
V&=&\frac{2q_V}{(r+Q)}dt-2p_V\cos\theta d\varphi\, ,\\ \label{eq:solzero}
\phi&=&\phi_{\infty}\, ,\quad c=c_{\infty}\, ,\quad H=0\, .
\end{eqnarray}
where $q_{A,V}$, $p_{A,V}$ are the electric and magnetic charges of the vectors $A$ and $V$ (in Planck units) and $Q=\sqrt{q_A^2+p_A^2+q_V^2+p_V^2}$ is the total duality-invariant charge.\footnote{Invariant under electric-magnetic duality and T-duality.}  On the other hand, the mass of this black hole is $M=Q$.  It is easy to check that this is a solution of (\ref{eq:hetred}) if the charges satisfy the following conditions
\begin{align} \nonumber
|q_A|=|p_A|\, ,\quad |q_V|=|p_V|&\, ,\\
q_{A}p_{V}+p_{A}q_{V}=0&\, .
\end{align}
The first two conditions ensure that $F^2=G^2=0$ while the third one implies that $F\wedge G=0$, and in this way the scalar fields have no sources. This special point in charge space has the property that the scalars are trivial at the supergravity level, although this does not hold once higher-curvature corrections are implemented, as we show below. Let us note that starting from a given solution, the transformation $t\rightarrow -t$ generates a new solution with opposite values of the electric charges and in turn $\phi\rightarrow-\phi$ changes the sign of the magnetic charges. Thus, without loss of generality we can consider $q_A=p_A>0$, and in that case $p_V=-q_V$. Hence,  there are two inequivalent sets of solutions, corresponding to $q_A\cdot q_V>0$ and $q_A\cdot q_V<0$.  We wish to compute the first-order $\alpha'$-corrections to these solutions, but for simplicity we will restrict to the case in which the absolute value of all the charges is the same. Hence, there are two possibilities: $q_A=q_V=p_A=-p_V=Q/2$ and $q_A=-q_V=p_A=p_V=Q/2$. In addition, the special case with $q_V=0$, in which $\alpha'$ corrections seem to introduce pathologies in the extremal limit, is treated in Appendix \ref{sec:case3}.

Thus, in this article we study the corrections to a stringy Reissner-Nordstr\"om black hole which, despite having two independent $U(1)$ dyonic vector fields, has only one independent charge $Q$. The configuration \eqref{eq:solzero} has an event horizon at $r=0$, with near-horizon geometry $AdS_2 \times S^2$ and it is therefore a black hole with vanishing temperature. On the other hand, the configuration does not preserve any supersymmetry, this being related to the presence of dyonic vectors, as in Ref.~\cite{Khuri:1995xq}.

\subsection{Case 1:  $q_A\cdot q_V<0$}\label{sec:case1}
Let us first consider the case $q_A=-q_V=p_A=p_V=Q/2$. Starting from the zeroth-order solution \eqref{eq:solzero} and using (\ref{eq:comapct}), it is possible to compute the first higher-curvature corrections by solving perturbatively the ten-dimensional equations of motion at first order in $\alpha'$, which can be found in Appendix \ref{sec:theory}. The details about the resolution of those equations are shown in Appendix \ref{sec-eom}. We find the following solution
\begin{eqnarray}
\notag
ds^2&=&\left(1+\frac{Q}{r}+\frac{\alpha'Q^2}{8(r+Q)^3r}\right)^{-2}dt^2-\left(1+\frac{Q}{r}+\frac{\alpha'Q^2}{8(r+Q)^3r}\right)^2\left(dr^2+r^2d\Omega_{(2)}^2\right)\, ,\\\notag
F&=&\frac{Q}{(r+Q)^2}\left(1+\frac{\alpha'Q^2}{4(r+Q)^4}\right)dt\wedge dr+Q\left(1+\frac{\alpha'Q(Q+4r)}{2(r+Q)^4}\right) \sin\theta d\theta\wedge d\varphi\, ,\\ \notag
V&=&-\frac{Q}{(r+Q)}dt-Q\cos\theta d\varphi\, ,\\\notag
\hat\phi&=&\hat\phi_{\infty}+\frac{\alpha'Q^2}{4(r+Q)^4}\, ,\\ \label{eq:sol1}
c&=&c_{\infty}\left(1+\frac{\alpha'Q^2}{4(r+Q)^4}\right)\, ,\quad H=0\, .
\end{eqnarray}

The conditions that we have imposed in order to solve the equations of motion are the same than those imposed for the original supergravity solution, namely
\begin{itemize}
\item regularity of the event horizon located at $r=0$,
\item fixed asymptotic value of the scalars: $\hat\phi\rightarrow\hat\phi_{\infty}$ and $c\rightarrow c_{\infty}$,
\item the metric is asymptotically flat, with the correct normalization at infinity, and
\item absence of additional \emph{free} charges at order $\alpha'$.
\end{itemize}

The last point means that we do not introduce artificial shifts in the charges. In fact, performing a transformation of the form $Q\rightarrow Q+\alpha' \delta Q$ in the original solution generates a new solution which, apparently, contains $\alpha'$-corrections. The integration constants of the equations of motion have to be appropriately chosen so that this type of shift does not occur. 

Observe that in the previous solution, $V$ contains no corrections. Also, note that $F$ is not a closed form, $dF\neq0$, so its local expression is no longer given by the exterior derivative of the vector field $A$. This is due to the form of the decomposition (\ref{eq:comapct}) and to the fact that $H$ is not a closed 3-form at first order in $\alpha'$. Thus, $F$ will have an expression of the form $F=dA+\alpha' W$ for some 2-form $W$. 
Nevertheless, the correct identification of the charges carried by these vector fields can be expressed in terms of $F$ and $G$ as follows\footnote{These integral expressions for the charges are valid in the asymptotic sphere $\mathbb{S}^2_{\infty}$. At a generic sphere some of the expressions contain additional higher-derivative terms, which vanish asymptotically.}$^,$\footnote{In the case of the charges associated with $F$, these can be written in terms of the ten-dimensional Kalb-Ramond field strength as follows,
\begin{eqnarray} \notag
q_A &=&=\frac{g_s^2}{8\pi (2\pi\ell_s)^5}\int_{\mathbb{S}^2_{\infty}\times \mathbb{T}^5} e^{-2\hat\phi}\star\hat H
\, ,\\ \notag
p_A&=&\frac{1}{8\pi c_{\infty}(2\pi\ell_s)}\int_{\mathbb{S}^2_{\infty}\times \mathbb{S}^1} \hat H\, .
\end{eqnarray}
}
\begin{eqnarray} \notag
q_A &=&\frac{1}{8\pi}\int_{\mathbb{S}^2_{\infty}}\frac{c^2_{\infty}}{c^2}e^{-2(\phi-\phi_{\infty})}\star F
\, ,\\ \notag
p_A&=&\frac{1}{8\pi}\int_{\mathbb{S}^2_{\infty}} F
\, ,\\ \notag
q_V&=&\frac{1}{8\pi}\int_{\mathbb{S}^2_{\infty}}e^{-2(\phi-\phi_{\infty})}\star G\, ,\\
p_V&=&\frac{1}{8\pi}\int_{\mathbb{S}^2_{\infty}} G\, .
\end{eqnarray}
We have checked that the evaluation of the integrals yields $q_A=p_A=-q_V=p_V=Q/2$, so that the charges of the solution are indeed unmodified and $Q$ is the total charge. One might think that corrections to the charges should not be expected, as they are defined by asymptotic integrals, where the curvature goes to zero. However, there are many examples of solutions for which higher-curvature interactions behave as delocalized sources of charge \cite{Cano:2018qev,Cano:2018brq,Cano:2018hut, Faedo:2019xii}, and hence it is always convenient to perform this computation.

The ADM mass of the black hole can be read from the asymptotic expansion of the metric according to
\begin{equation}
\lim_{r \rightarrow \infty} g_{rr}=1+\frac{2M}{r}+\ldots\, .
\end{equation}
From \eqref{eq:sol1} on sees that $M=Q$, so the charge-to-mass ratio of this extremal black hole is not modified at this order,

\begin{equation}
\frac{Q}{M}=1+\mathcal{O}(\alpha'^2)\, .
\end{equation}

It is also interesting to compute the correction to the entropy of this black hole. The application of Wald's formula for this family of solutions of the heterotic theory is described in Appendix \ref{sec:wald}. Upon evaluation of the resulting expression \eqref{eq:wald2} for the background \eqref{eq:sol1} we get
\begin{equation}
\mathbb{S}=\frac{\pi}{G_{N}^{(4)}} \left(Q^2+\frac{\alpha'}{4} \right) \, .
\end{equation}
\noindent
The first term in the expression corresponds to the Bekenstein-Hawking entropy. Therefore, we find a positive correction of the entropy, that can be interpreted as capturing additional microscopic degrees of freedom that are frozen in the truncation to the two-derivative supergravity theory. At the computational level, this deviation is originated from an increase in the area of the event horizon, while the contributions in Wald's formula coming explicitly from the higher-derivative terms vanish.

\subsection{Case 2:  $q_A\cdot q_V>0$}\label{sec:case2}
Let us now consider the case with $q_A=q_V=p_A=-p_V=Q/2$. The first-order in $\alpha'$ corrections turn out to be quite different. The solution reads
\begin{eqnarray} \notag
ds^2&=&A^2\left(1+\frac{Q}{r}\right)^{-2}dt^2-B^2\left(1+\frac{Q}{r}\right)^2\left(dr^2+r^2d\Omega_{(2)}^2\right)\, ,\\\notag
F&=&\frac{Q}{(r+Q)^2}\left(1+\alpha'\frac{3 Q^2-10 Q r-3 r^2}{120 (r+Q)^4}\right)dt\wedge dr+Q\left(1-\frac{\alpha' Q^2}{2 (r+Q)^4}\right) \sin\theta d\theta\wedge d\varphi\, ,\\\notag
V&=&\frac{Q}{(r+Q)}\left(1-\alpha'\frac{r (63 Q+r)}{120 (r+Q)^4}\right)dt+Q\cos\theta d\varphi\, ,\\\notag
\hat\phi&=&\hat\phi_{\infty}-\frac{\alpha' r \left(19 Q^2+12 Q r+3 r^2\right)}{60 Q (r+Q)^4}\, ,\\
c&=&c_{\infty}\left(1-\frac{\alpha'Q^2}{4(r+Q)^4}\right)\, ,\quad H=0\, ,
\end{eqnarray}
where

\begin{align}
A=&1+\alpha'\frac{6 Q^3+13 Q^2 r+8 Q r^2+2r^3}{40 Q (r+Q)^4}\, ,\\
B=&1-\alpha' \frac{5 Q^3+9 Q^2 r+7 Q r^2+2 r^3}{40 Q (r+Q)^4}\, .
\end{align}

In order to obtain this solution we have imposed the same conditions as in the previous case.
Notice that now $V$ receives corrections, while $F$ is again not closed. Still, one can check that the charges have the correct values $q_A=p_A=q_V=-p_V=Q/2$, and therefore, $Q$ is indeed the total charge. We also note that, unlike in the previous solution, the dilaton acquires a non-trivial charge that cannot be removed, and it reads\footnote{It is identified by the asymptotic expansion $\hat\phi=\hat\phi_{\infty}-Q_{\hat\phi}/r+\mathcal{O}(1/r^2)$.}
\begin{equation}
Q_{\hat\phi}=\frac{\alpha'}{20Q} \, .
\end{equation}

On the other hand, the metric component $g_{rr}$ behaves asymptotically as
\begin{equation}
g_{rr}=1+\frac{1}{r}\left(2Q-\frac{\alpha'}{10 Q}\right)+\ldots\, .
\end{equation}
and therefore, the mass receives corrections in this case
\begin{equation}
M=Q-\frac{\alpha'}{20Q}+\mathcal{O}(\alpha'^2)\, .
\end{equation}
Correspondingly, the extremal charge-to-mass ratio is modified
\begin{equation}
\frac{Q}{M}=1+\frac{\alpha'}{20 M^2}+\mathcal{O}(\alpha'^2)\, ,
\end{equation}
and it is larger than one, in agreement with the mild form of the WGC. 

The Wald entropy of this solution is obtained from the evaluation of \eqref{eq:wald2}. In this case there is a negative contribution from the modification of the area of the event horizon as well as a positive one from the higher-derivative terms. The result reads
\begin{equation}
\mathbb{S}=\frac{\pi}{G_{N}^{(4)}} \left(Q^2+\frac{\alpha'}{4} \right) \, ,
\end{equation}
which surprisingly enough, coincides with the value found in the previous case, even though the rest of the properties of the solution are different.

\section{Discussion}
\label{sec-discussion}

In this article we have analyzed the effect produced by higher-curvature
corrections to a family of extremal, non-supersymmetric black holes in the
context of heterotic superstring theory. We have found the first example of
modification of the charge-to-mass ratio of an extremal black hole explicitly
embedded in string theory. This example defies previous expectations that the
charge-to-mass ratio of extremal black holes in a supersymmetric theory is not
modified by higher-curvature corrections. Likewise, we have presented evidence
that such modifications are not necessarily in correspondence with the
corrections to the entropy of the black hole. While this differs from the
result of the explorations performed in Einstein-Maxwell theory, it agrees
with earlier results on the supersymmetric three- and four-charge systems of
the heterotic theory, as we mentioned in the introduction.  The difference with the results in Refs.~\cite{Cheung:2018cwt,Cheung:2019cwi,Goon:2019faz} can
be understood if one takes into account two facts. First of all, topological terms such as 
the Gauss-Bonnet invariant --- which implicitly appears in the heterotic string effective action --- modify
the black hole entropy while keeping the solution unchanged. Thus, in this case corrections to the entropy are independent
from deviations of the extremal charge-to-mass ratio. On the other hand, the models considered in the previous literature 
do not include scalars, which are a key ingredient of stringy effective actions. As we have seen in the examples presented,
these scalars are activated by higher-derivative corrections even if they are trivial in the zeroth-order solution. Scalar fields usually affect the thermodynamic description of black holes --- see \textit{e.g.} \cite{Astefanesei:2018vga} --- and it would be
interesting to explore whether this could modify the conclusions of \cite{Goon:2019faz}.

The string coupling constant and the curvature can be kept sufficiently small in the exterior region of the black hole for the cases we have considered, hence the low energy field theory description gives a good approximate description of the system. In our analysis, we focused on the case $|q_A|=|q_V|$ for simplicity, but it would be interesting to study the corrections to the solution (\ref{eq:solzero}) for general values of $q_A$ and $q_V$, so that the two dyonic vectors are independent. In that case, we expect that the extremality bound will be modified according to
\begin{equation}
\frac{Q}{M}\le1+\frac{\alpha'}{M^2}f(q_A,q_V)+\mathcal{O}(\alpha'^2)\, ,
\end{equation}
where $f$ is a certain homogeneous function of degree 0 of the charges $q_A$ and $q_V$, which, according to the WGC, should be non-negative.  
In this paper, we have shown that
\begin{equation}
f(q,-q)=0\, ,\quad f(q,q)=\frac{1}{20}\, .
\end{equation}
In addition, in Appendix \ref{sec:case3} we consider the case $q_V=0$ and we show that \linebreak$f(q,0)=\frac{1}{80}$.\footnote{The study of the non-extremal black holes of that type will be the object of a coming publication \cite{RNCORR}.  $\alpha'$-corrections seem to introduce pathologies in the extremal case even though the zeroth-order solution is regular.} Furthermore, since the result must be invariant under T-duality we conclude that $f(0,q)=f(q,0)$.
Given the values found, it is an interesting problem to search for the general expression of the function $f(q_A,q_V)$ and to check its non-negativity.

An important piece in our analysis is that the higher-curvature corrections to the theory are directly taken from the ten-dimensional heterotic string theory. This differs from the approach usually taken in the literature, where given a four-dimensional effective theory, all possible four-derivative terms that can be constructed with the corresponding field content are considered. As we have mentioned in the main text, consistency may require to enlarge the field content of an effective theory when perturbative corrections are being considered. Of course, the details depend on the UV theory on which the system is embedded. It is interesting to mention here the example of the dyonic Reissner-Nordstrom black hole solution of Einstein-Maxwell-Dilaton theory embedded in the heterotic theory ($f(q,0)$ configuration), which requires the activation of additional fields that can be truncated at the two-derivative level \cite{RNCORR}.

The mild version of the WGC affects solutions well above Planck mass. Hence, it has little predictive power about low energy effective theories in itself. It has been recently suggested that it might be possible to relate the mild and strong versions of the WGC using modular invariance of string theory \cite{Aalsma:2019ryi}. Given the growing amount of evidence in favor of the mild WGC, this seems a promising idea and it would be interesting to test it for a regular extremal black hole system of string theory.\footnote{Reference \cite{Aalsma:2019ryi} considered two-charge black holes, which have singular horizon in the extremal limit even after higher-curvature corrections are included \cite{Cano:2018hut}.}

\section*{Acknowledgments}

This work has been supported in part by the INFN, the MCIU, AEI, FEDER (UE) grant
PGC2018-095205-B-I00 and by the Spanish Research Agency (Agencia Estatal de
Investigaci\'on) through the grant IFT Centro de Excelencia Severo Ochoa
SEV-2016-0597. The work of PAC is funded by Fundaci\'on la Caixa through a ``la Caixa - Severo Ochoa'' International pre-doctoral grant. PFR would like thank the Albert Einstein Institute at Potsdam
for hospitality while this work was being completed.  TO wishes to thank
M.M.~Fern\'andez for her permanent support.

\appendix

\section{The heterotic theory}
\label{sec:theory}

In this article we work in the context of the heterotic superstring effective action at first order in the $\alpha'$ expansion as constructed in \cite{Bergshoeff:1989de}. We consider black hole solutions with small string coupling $g_s$ and sufficiently large horizon such that the supergravity approximation is valid at least outside the event horizon. Still, we are interested in looking at the information that the dominant corrections of higher order in curvature produce in some properties of the solutions.  The effective action of the heterotic superstring at first order in $\alpha'$ is given by

\begin{equation}
\label{action}
{S}
=
\frac{g_{s}^{2}}{16\pi G_{N}^{(10)}}
\int d^{10}x\sqrt{|{\hat g}|}\, 
e^{-2{\hat \phi}}\, 
\left\{
\hat{R} 
-4(\partial{\hat \phi})^{2}
+\frac{1}{2\cdot 3!}{\hat H}^{2}
-\frac{\alpha'}{8} \hat R_{(-)}{}_{\mu\nu}{}^a{}_b \hat R_{(-)}{}^{\mu\nu\, b}{}_a +\dots
\right\}\, .
\end{equation}
We use hats to denote the ten-dimensional heterotic fields. We have not included Yang-Mills fields in the theory for simplicity\footnote{Some examples with non-trivial Yang-Mills fields were given in \cite{Cano:2018qev, Chimento:2018kop, Cano:2018brq}.}. The curvature of the torsionful spin connection, defined as $\hat \omega_{(-)}{}^a{}_b=\hat\omega^a{}_b-\frac{1}{2}\hat H_\mu{}^a{}_b \, dx^\mu$, is 
\begin{equation}
\hat R_{(-)}{}^a{}_b = d\hat \omega_{(-)}{}^a{}_b - \hat \omega_{(-)}{}^a{}_c \wedge \hat \omega_{(-)}{}^c{}_b \, .
\end{equation}

\noindent
The field strength $H$ of the Kalb-Ramond 2-form $B$ includes a Chern-Simons term 
\begin{equation}
\label{def:kalb-ramond}
\hat H = d\hat B+ \frac{\alpha'}{4} \hat \Omega^{\text{L}}_{(-)} \,,
\end{equation}
where
\begin{equation}
\label{def:chern-simons}
\hat \Omega^{\text{L}}_{(-)} = d\hat \omega_{(-)}{}^a{}_b \wedge \hat \omega_{(-)}{}^b{}_a - \frac{2}{3} \hat\omega_{(-)}{}^a{}_b \wedge \hat\omega_{(-)}{}^b{}_c \wedge \hat\omega_{(-)}{}^c{}_a \,.
\end{equation}
Then, the corresponding Bianchi identity reads
\begin{equation}\label{eq:bianchi}
d\hat H=\frac{\alpha'}{4}\hat R_{(-)}{}^a{}_b\wedge \hat R_{(-)}{}^b{}_a \, .
\end{equation}

\noindent
The equations of motion derived from the action \eqref{action} are 
\begin{eqnarray}
\label{eq:eq1}
\hat R_{\mu\nu} -2\nabla_{\mu}\partial_{\nu}\hat\phi
+\frac{1}{4}\hat{H}_{\mu\rho\sigma}\hat{H}_{\nu}{}^{\rho\sigma}
-\frac{\alpha'}{4}\hat R_{(-)}{}_{\mu\rho}{}^a{}_b \hat R_{(-)}{}_\nu{}^{\rho\, b}{}_a
& = & 
\mathcal{O}(\alpha'^2)\, ,
\\
& & \nonumber \\
\label{eq:eq2}
(\partial \hat \phi)^{2} -\frac{1}{2}\nabla^{2}\hat \phi
-\frac{1}{4\cdot 3!}\hat{H}^{2}
+\frac{\alpha'}{32}\hat R_{(-)}{}_{\mu\nu}{}^a{}_b \hat R_{(-)}{}^{\mu\nu\, b}{}_a
& = &
\mathcal{O}(\alpha'^2)\, ,
\\
& & \nonumber \\
\label{eq:eq3}
d\left(e^{-2\hat\phi}\star\!\hat{H}\right)
& = &
\mathcal{O}(\alpha'^2)\, .
\end{eqnarray}

\noindent

The (zeroth-order) supergravity theory can be recovered from these expressions
by setting
$\alpha'=0$. 
The action includes a tower of corrections of all powers in $\alpha'$ due to
the recursive definition of the Kalb-Ramond field strength, which breaks the
supersymmetry of the supergravity theory. The term of quadratic order in
curvature at \eqref{action} was found imposing supersymmetry of the theory at
first order in $\alpha'$ after inclusion of the Chern-Simons term in the field
strength \cite{Bergshoeff:1988nn}. Further corrections of higher power in the
curvature $R_{(-)}$ of the torsionful spin connection are required to recover
supersymmetry order by order. Additional higher-curvature corrections
unrelated to the supersymmetrization of the Kalb-Ramond kinetic term are also
present, although those appear first at cubic order in
$\alpha'$. 

\section{Solving the equations of motion}
\label{sec-eom}

We consider the ansatz for the 10-dimensional fields introduced in Eq.~(\ref{eq:comapct}),
\begin{align} \nonumber 
d\hat{s}=&\frac{c_{\infty}}{c}e^{2( \hat\phi- \hat\phi_{\infty})}ds^2-c^2(dz+V/c_{\infty})^2-dy^idy^i\, , \\ \nonumber
\hat H=&F\wedge (c_{\infty}dz+V)+H\, ,
\end{align}
and we assume the following form of the building blocks
\begin{align}
ds^2&=(1+\alpha' a)^2\left(1+\frac{Q}{r}\right)^{-2}dt^2-(1+\alpha' b)^2\left(1+\frac{Q}{r}\right)^2\left(dr^2+r^2d\Omega_{(2)}^2\right)\ ,\\
F&=\frac{Q}{(r+Q)^2}\left(1+\alpha' d(r)\right)dt\wedge dr+Q\left(1+\alpha' e(r)\right) \sin\theta d\theta\wedge d\varphi\, ,\\
V&=-\frac{\epsilon Q}{(r+Q)}\left(1+\alpha' f(r)\right)dt-\epsilon Q\cos\theta d\varphi\, \\
H&=\alpha' g(r)\sin\theta dt\wedge d\theta\wedge d\varphi\, ,\\
\hat\phi&=\hat\phi_{\infty}+\alpha'\delta\hat\phi(r)\, ,\\
c&=c_{\infty}(1+\alpha'\delta c(r))\, ,
\end{align}
where $Q>0$ and $\epsilon=\pm 1$ is a sign. For $\epsilon=+1$ and $\epsilon=-1$ we have, respectively, the cases 1 and 2 discussed in Sections~\ref{sec:case1} and \ref{sec:case2}. In the limit $\alpha'\rightarrow 0$, this ansatz reduces to the zeroth-order solution in Eq.~(\ref{eq:sol1}). The $\alpha'$ corrections are introduced through the functions $a$, $b$, $\delta c$, $d$, $e$, $f$, $g$ and $\delta\phi$, and one can see that this ansatz is general enough in order to solve the first-order-in-$\alpha'$ equations.

On the one hand, keeping only the terms up to linear order in $\alpha'$, the Bianchi identity (\ref{eq:bianchi}) reads 
{\allowdisplaybreaks
\begin{align}
\notag
 &d\hat H-\frac{\alpha'}{4}\hat R_{(-)}{}^a{}_b\wedge \hat R_{(-)}{}^b{}_a= \alpha'  \sin \theta  \left[dt\wedge dr\wedge d\theta\wedge d\varphi\left(\frac{d Q^2 \epsilon }{(Q+r)^2}\right.\right.\\&\notag\left.\left.+\frac{Q^3 (Q-Q \epsilon +r (5+3 \epsilon
   ))}{(Q+r)^6}+\frac{Q^2 \epsilon  e'}{Q+r}-\frac{Q^2 \epsilon  \left(e+f-(Q+r)f'\right)}{(Q+r)^2}-g'\right)\right.\\&\left.+Q c_{\infty } dr\wedge d\theta\wedge d\varphi \wedge dz \left(\frac{Q (Q (-1+\epsilon )+3 r (1+\epsilon ))}{(Q+r)^5}+e'\right) \right]
\end{align}
On the other hand, the equations (\ref{eq:eq3}), (\ref{eq:eq2}) and the relevant components of Einstein's equations (\ref{eq:eq1}) yield
\begin{align}
&d \left(e^{-2\hat\phi}\star\hat H\right)=-Q \alpha'  \sin \theta dr\wedge d\theta\wedge d\varphi \left(\frac{g \epsilon}{r^2} +
   \left(a'-b'-d'+\delta c'+2 \delta \phi '\right)\right)\, ,\\
   \mathcal{E}_{\hat\phi}&=\alpha'  \left(Q^2 \left(\frac{Q^2-12 r^2}{8(Q+r)^8}+\frac{ (a-b-d+e) }{2(Q+r)^4}\right)+\frac{(r^2\delta \phi ')'}{2(Q+r)^2} \right)\, ,\\
   \mathcal{E}_{tt}&=\frac{\alpha'}{(Q+r)^4}\left[\frac{e Q^4}{(Q+r)^2}-\frac{2 a Q^2 r^2}{(Q+r)^2}+\frac{d Q^2 (-Q+r)}{Q+r}+\frac{f Q^2
   \left(Q^2+r^2\right)}{(Q+r)^2}+\frac{Q^2 r^2 \text{$\delta $c}}{(Q+r)^2}\notag\right.\\&\left.-\frac{2 Q^2
   r^2 \delta \phi }{(Q+r)^2}-\frac{g Q^2 \epsilon }{Q+r}-\frac{Q^2 r^2 \left(12 Q r
   (-1+\epsilon )+6 r^2 (-1+\epsilon )+Q^2 (5+17 \epsilon )\right)}{4
   (Q+r)^6}\notag\right.\\&\left.+\frac{r^2 \left(Q^2-3 Q r-2 r^2\right) a'}{Q+r}-Q r^2 b'-\frac{Q^2
   \left(Q^2+r^2\right) f'}{Q+r}+\frac{r (2 Q+r) \left(Q^2-Q r+r^2\right) \text{$\delta
   $c}'}{Q+r}\notag\right.\\&\left.-\frac{2 r^2 \left(-Q^2+Q r+r^2\right) \delta \phi '}{Q+r}-r^4 a''+Q^2 r^2
   f''+\left(Q^2 r^2+\frac{r^4}{2}\right) \text{$\delta $c}''-r^4 \delta \phi ''\right]\, ,\\
   \mathcal{E}_{rr}&=\alpha'\left[\frac{2 a Q^2}{r^2 (Q+r)^2}-\frac{d Q^2}{r^2 (Q+r)^2}-\frac{f Q^2}{r^2
   (Q+r)^2}-\frac{Q^2 \text{$\delta $c}}{r^2 (Q+r)^2}+\frac{2 Q^2 \delta \phi }{r^2
   (Q+r)^2}
   \notag\right.\\&\left.+\frac{Q^4 (1+\epsilon )-2 Q^2 r^2 (5+\epsilon )}{4 r^2 (Q+r)^6}+\frac{3 Q
   a'}{r (Q+r)}+\frac{(Q+2 r) b'}{Q r+r^2}+\frac{Q^2 f'}{r^2 (Q+r)}\notag\right.\\&\left.-\frac{\text{$\delta
   $c}'}{r}+\frac{2 \delta \phi '}{r}+a''+2 b''-\frac{\text{$\delta $c}''}{2}+\delta
   \phi ''\right]\, ,\\
   \mathcal{E}_{zz}&=\frac{\alpha'c_{\infty}^2}{(Q+r)^2}\left[-\frac{d Q^2}{(Q+r)^2}+\frac{e Q^2}{(Q+r)^2}+\frac{f Q^2}{(Q+r)^2}-\frac{3 Q^2 r^2
   \epsilon }{(Q+r)^6}-\frac{Q^2 f'}{Q+r}+2 r \text{$\delta $c}'+r^2 \text{$\delta
   $c}''\right]\, ,\\
   \mathcal{E}_{tz}&=\frac{\alpha'c_{\infty}}{(Q+r)^3}\left[\frac{g Q}{2 (Q+r)}+\frac{d Q^3 \epsilon }{(Q+r)^2}-\frac{e Q^3 \epsilon
   }{(Q+r)^2}-\frac{f Q^3 \epsilon }{(Q+r)^2}\notag\right.\\&\left.+\frac{Q^2 r^2 (-3 r (-1+\epsilon )+Q
   (7+\epsilon ))}{2 (Q+r)^6}-\frac{Q r^2 \epsilon  a'}{2 (Q+r)}+\frac{Q r^2 \epsilon 
   b'}{2 (Q+r)}+\frac{Q^3 \epsilon  f'}{Q+r}\notag\right.\\&\left.-\frac{Q r (4 Q+r) \epsilon  \text{$\delta
   $c}'}{2 (Q+r)}-\frac{Q r^2 \epsilon  \delta \phi '}{Q+r}-\frac{1}{2} Q r^2 \epsilon 
   f''-Q r^2 \epsilon  \text{$\delta $c}''\right]\, ,\\
   \mathcal{E}_{\varphi z}&=\frac{\alpha'c_{\infty}\cos\theta}{(Q+r)^2}\left[\frac{3 Q^3 r^2}{(Q+r)^6}+\frac{d Q^3 \epsilon }{(Q+r)^2}-\frac{e Q^3 \epsilon
   }{(Q+r)^2}-\frac{f Q^3 \epsilon }{(Q+r)^2}+\frac{Q^3 \epsilon  f'}{Q+r}-2 Q r
   \epsilon  \text{$\delta $c}'-Q r^2 \epsilon  \text{$\delta $c}''\right]
\end{align}
}

Note that here we have written nine equations for eight variables, but of course not all of them are independent, since they are related by Bianchi identities. Nevertheless, it is convenient to use all of the equations above in order to simplify the problem.   By combining the equations in an appropriate way it is possible to find the general solution, which contains a large number of integration constants. These constants are then fixed by the conditions specified in the main text, which, more precisely, can be expressed as follows:
\begin{itemize}
\item Regularity at $r=0$: all of the functions $a$, $b$, $\delta c$, $d$, $e$, $f$, $g$, $\delta\phi$ are finite at $r=0$. 
\item Fixed asymptotic values of the scalars: $\delta\hat\phi(r)\rightarrow 0$, $\delta c(r)\rightarrow 0$, $g(r)\rightarrow 0$ when $r\rightarrow \infty$.
\item Asymptotic flatness: $a(r)\rightarrow 0$, $b(r)\rightarrow 0$ when  $r\rightarrow \infty$.
\item Absence of additional free charges at order $\alpha'$:  $d(r)\rightarrow 0$, $e(r)\rightarrow 0$, $f(r)\rightarrow 0$ when  $r\rightarrow \infty$.
\end{itemize}
These conditions completely determine the solution and one finds

\begin{align}
a&=\frac{Q^3 (1-11\epsilon)+(1 -\epsilon)(13 Q^2 r +8 Q r^2 +2 r^3)}{80 Q (Q+r)^4  }\, \\
b&=\frac{10 \epsilon Q^3+r \left(9 Q^2+7 Q r+2 r^2\right) (\epsilon-1)}{80 Q (Q+r)^4 }\, ,\\
\delta c&= \frac{\epsilon Q^2}{4 (Q+r)^4  }\, ,\\
d&=\frac{(\epsilon-1 )(10 Q r +3 r^2 )+Q^2 (57\epsilon+63  )}{240 (Q+r)^4
    }\, ,\\
e&=\frac{Q (Q \epsilon +2 r (1+\epsilon ))}{2 (Q+r)^4}\, ,\\
f&=\frac{r (63 Q+r) (\epsilon-1)}{240 (Q+r)^4}\, .\\
g&=0\, ,   \\
\delta\hat\phi&=\frac{r \left(19 Q^2+12 Q r+3 r^2\right) (\epsilon-1)+15 Q^3 (\epsilon+1 )}{120 Q(Q+r)^4}\, .
\end{align}
For $\epsilon=\pm1$ one gets the results shown in Sections~\ref{sec:case1} and \ref{sec:case2}.

\section{Wald entropy}
\label{sec:wald}

The Wald entropy formula of the ten-dimensional theory is

\begin{equation}
\label{eq:wald}
 \mathbb{S}
=
-2\pi \int_\Sigma d^8x\sqrt{|h|}\mathcal{E}^{abcd}\epsilon_{ab}\epsilon_{cd}\,,
\end{equation}

\noindent
where $\Sigma$ is a cross-section of the horizon with induced metric $h_{\mu\nu}$, $\epsilon_{ab}$ is the binormal to $\Sigma$
normalized as $\epsilon_{ab}\epsilon^{ab}=-2$ and $\mathcal{E}^{abcd}$ is defined as

\begin{equation}
\mathcal{E}^{abcd} 
= \frac{g_s^2}{16\pi G_N^{(10)}}
\frac{\delta \mathcal{L}}{\delta \hat R_{abcd}}\, ,
\end{equation}

\noindent
where $\mathcal{L}$ is the Lagrangian of the theory \eqref{action}. We can apply this formula to the action of the heterotic theory for the family of regular solutions we study in this article. In order to compute the integrand it is convenient to use flat indices. We define the vielbein

\begin{gather} 
\nonumber
 e^{0} =e^{{\phi-\phi_\infty}}  e^{G}  dt\,,\qquad 
 e^{1}=  e^{\phi-\phi_\infty} e^{-U}  dr\,,\qquad  
 \\ \nonumber \\ \nonumber
 e^{2} = e^{\phi-\phi_\infty} e^{-U} r d\theta\,, \qquad 
 e^{3} = e^{\phi-\phi_\infty} e^{-U}r \sin\theta d\varphi\, ,
\\
\nonumber\\ 
e^{4} =  c \left( dz+\frac{V}{c_\infty} \right)\, , \qquad
e^i=dy^i \, .
\end{gather}

\noindent
Here $e^{2G}=g_{tt}$ and $e^{-2U}=g_{rr}$, with $g_{\mu\nu}$ the four-dimensional metric in the Einstein frame. The non-vanishing components of the binormal in flat indices are $\epsilon_{01}=-\epsilon_{10}=1$. The volume form entering Wald's formula is

\begin{equation}
d^8 x \sqrt{|h|}
= d\theta d\varphi dz d^4y
c_\infty e^{2(\hat \phi- \hat \phi_\infty)} e^{-2U} r^{2} \sin{\theta}   \,.
\end{equation}

The variation of the Lagrangian with respect to the Riemann tensor contains three non-vanishing contributions. The first one comes from the Einstein-Hilbert term in \eqref{action}, which amounts to

\begin{equation}
\mathcal{E}_0^{abcd}= \frac{e^{-2(\hat \phi-\hat \phi_\infty)}}{16\pi G_N^{(10)}}\frac{ \delta \hat R}{ \delta \hat R_{abcd}}= \frac{e^{-2(\hat\phi-\hat\phi_\infty)}}{16\pi G_N^{(10)}} \eta^{ac} \eta^{bd} \, ,
\end{equation}

\noindent
where $\eta^{ab}$ is the inverse flat metric. This term is responsible for the Bekenstein-Hawking leading order term in the entropy, $\mathbb{S}_0=A_{\Sigma}/4G_N^{(10)}$. Following \cite{Faedo:2019xii}, one can see that there are two additional contributions arising from the variation of the Chern-Simons 3-form in the Kalb-Ramond field strength.  Each of those is coming from one of the two factors in the decomposition 

\begin{equation}
\label{eq:decomCS}
\hat \Omega^{\text{L}}_{(-)} = \mathscr{A} +\hat \Omega^{\text{L}} \,,
\qe
where $\hat\Omega^{\text{L}}$ is the standard Lorentz Chern-Simons term in terms of the spin connection $\hat\omega^a_{\ b}$, and
\eq
\begin{aligned}
\mathscr{A} = \frac12 \, d(\hat\omega^a_{\ b} \wedge \hat H^b_{\ a}) + \frac14 \hat H^a_{\ b} \wedge D\hat H^b_{\ a} - \hat R^a_{\ b} \wedge \hat H^b_{\ a} + 
\frac{1}{12} \hat H^a_{\ b} \wedge\hat H^b_{\ c} \wedge\hat H^c_{\ a} \, .
\end{aligned}
\qe
Here $\hat H^a_{\ b} =\hat H_{\mu\ b}^{\ a} dx^\mu$ and $D\hat H{}^a_{\ b} = d\hat H{}^a_{\ b} +\hat \omega^a_{\ c} \wedge \hat H{}^c_{\ b} - \hat \omega^c_{\ b} \wedge \hat H{}^a_{\ c}$. Notice that the last term in \eqref{action} gives no contribution to the entropy, since it is quadratic in the curvature of the torsionful spin connection, which vanishes at the horizon. 

Using this rewriting, in first place we get

\begin{equation}
\mathcal{E}_1^{abcd}=\frac{e^{-2(\hat\phi-\hat\phi_\infty)}}{16\pi G_N^{(10)}} \frac{ \delta }{ \delta \hat R_{abcd}} \left( \frac{\alpha'}{3! 4 }\hat{H}^{efg} \mathscr{A}_{efg} \right)= \frac{e^{-2(\hat\phi-\hat\phi_\infty)}}{16\pi G_N^{(10)}} \frac{\alpha'}{8}\hat H^{abf} \hat H_f\,^{cd} \, .
\end{equation}

\noindent
To obtain the last correction to the entropy, we notice that when $\mathcal{E}^{abcd}$ gets contracted with the binormal, the only relevant values of the flat indices $a,\dots,d$ are $0,1$. Therefore, the remaining non-vanishing contribution to the entropy comes from the variation of $\hat\Omega^L_{014}$, and amounts to

\begin{equation}
\mathcal{E}_2^{abcd}=\frac{e^{-2(\hat\phi-\hat\phi_\infty)}}{16\pi G_N^{(10)}} \frac{ \delta }{ \delta \hat R_{abcd}} \left( \frac{\alpha'}{3! 4 }\hat{H}^{efg} \Omega^L_{efg} \right)= \frac{e^{-2(\hat\phi-\hat\phi_\infty)}}{16\pi G_N^{(10)}} \frac{\alpha'}{8}\hat H^{ab4}\frac{c}{c_\infty} G^{cd}  \, ,
\end{equation}

Putting everything together, Wald's entropy is
\begin{equation}
\label{eq:wald2}
\mathbb{S}
= 
\frac{1}{4 G_{N}^{(10)}}\int d\theta d\varphi dz d^4y c_\infty
e^{-2U} r^2  \sin{\theta}
 \left[1-\frac{\alpha'}{4} \hat H^{014}
   \left(\hat H_4\,^{01}+\frac{c}{c_\infty} {G}^{01} \right) \right]\,.
\end{equation}

\section{Case 3:  $q_A\cdot q_V=0$ }
\label{sec:case3}

As we have seen in the main text, the cases $q_A\cdot q_V>0$ and $q_A\cdot q_V<0$ are qualitatively very different when higher-curvature corrections are incorporated. 
It is interesting to consider the transition between one and another,  $q_A\cdot q_V=0$. In particular let us consider the configuration $q_V=p_V=0$, $q_A=p_A=Q/\sqrt{2}$. Then we find the following solution, 
\begin{eqnarray}
ds^2&=&A^2\left(1+\frac{Q}{\rho}\right)^{-2}dt^2-B^2\left(1+\frac{Q}{\rho}\right)^2\left(d\rho^2+\rho^2d\Omega_{(2)}^2\right)\, ,\\
F&=&\frac{c}{c_{\infty}}\frac{e^{2 (\hat\phi-\hat\phi_{\infty})}\sqrt{2}QA }{(\rho+Q)^2B}dt\wedge d\rho+\sqrt{2}Q E\sin\theta d\theta\wedge d\varphi\, ,\\
V&=&\frac{\alpha' \rho^2 \left(Q \left(11 Q^2+15 Q \rho+6 \rho^2\right)-6 (Q+\rho)^3 \log
   \left(1+\frac{Q}{\rho}\right)\right)}{3 \sqrt{2} Q^2 (Q+\rho)^5} dt\, ,\\
   \hat\phi&=&\hat\phi_{\infty}+\frac{\alpha'}{1120 Q^2 (Q+\rho)^4} \Bigg[Q \left(-189 Q^3-196 Q^2 \rho+1646 Q \rho^2+1304 \rho^3\right)\\
   &+&\nonumber 28 (Q+\rho)^2 \left(5 Q^2+10 Q\rho-47 \rho^2\right) \log \left(1+\frac{Q}{\rho}\right)\Bigg]\, ,\\
    c(\rho)&=&c_{\infty}+\frac{\alpha'}{560 Q^2 (Q+\rho)^4}\Bigg[Q \left(259 Q^3+896 Q^2 \rho+1614 Q \rho^2+876 \rho^3\right)\\ \nonumber
   &-&28 (Q+\rho)^2 \left(5 Q^2+10 Q
   \rho+31 \rho^2\right) \log \left(1+\frac{Q}{\rho}\right)\Bigg]\, ,\\
H&=& -\frac{\alpha'Q^2 \rho^2}{(Q+\rho)^5}dt\wedge \sin\theta d\theta\wedge d\varphi- F\wedge V\, ,
\end{eqnarray}
where 
\begin{eqnarray}
A&=&1+\alpha'\frac{83 Q^3+189 Q^2 \rho+84 Q \rho^2+6 \rho^3-60 (Q+\rho)^3 \log \left(1+\frac{Q}{\rho}\right)}{480 Q
(Q+\rho)^4}\, ,\\
B&=&1+\alpha'\frac{Q (5 Q+3 \rho) \left(6 Q^2+13 Q \rho+18 \rho^2\right)-60 \rho (Q+\rho)^3 \log
   \left(1+\frac{Q}{\rho}\right)}{480 Q^2 (Q+\rho)^4}\, ,\\
   E&=&1+\frac{\alpha' Q (Q+4 \rho)}{4 (Q+\rho)^4}\, .\\
  \end{eqnarray}

We observe that there are logarithmic singularities that cannot be avoided. 
Now, the charges of the solution are indeed $q_A=p_A=Q/\sqrt{2}$. Thus, $Q$ still represents the total charge $Q=\sqrt{q_A^2+p_A^2}$. 
In addition, we note that the KK vector field $V$ does not carry any charge. However, we get
\begin{equation}
g_{\rho\rho}=1+\frac{1}{\rho}\left(2Q-\frac{\alpha'}{40 Q}\right)+\ldots\, .
\end{equation}
Therefore, the mass is 
\begin{equation}
M=Q-\frac{\alpha'}{80Q}+\ldots\Rightarrow \frac{Q}{M}=1+\frac{\alpha'}{80 M^2}+\mathcal{O}(\alpha'^2)
\end{equation}

\renewcommand{\leftmark}{\MakeUppercase{Bibliography}}
\phantomsection
\bibliographystyle{JHEP}
\bibliography{references}
\label{biblio}

\end{document}